# ULTRA-STABLE 3D-PRINTED PRECISION VOLTAGE DIVIDER FOR CALIBRATIONS AND EXPERIMENTS

*Stephan Passon* [a], *Kristian König* [b,*], *Florian Schilling* [a], *Bernhard Maaß* [b,c], *Johann Meisner* [a], *Wilfried Nörtershäuser* [b]

[a] PTB (High voltage metrology, Physikalisch-Technische Bundesanstalt), Braunschweig, Germany, stephan.passon@ptb.de:
[b] Technische Universität Darmstadt (Institut für Kernphysik), Darmstadt, Germany,
[c] Argonne National Laboratory, Chicago, IL, USA

*Abstract* − This paper presents the concept of an ultra-stable, thermally independent precision voltage divider tailored for direct current (DC) voltages up to 60 kV. Key features of this voltage divider include minimal voltage dependence, excellent stability, and resistance to external temperature variations. The innovative approach involves its fabrication using 3D printing technology, allowing easy replication by project partners. This precision voltage divider leverages commercially available precision resistors, drawing upon successful outcomes from the FutureEnergy 19ENG02 and HVDC ENG07 Projects.

In these experiments, which involve ion acceleration and laser probing of electronic transitions, voltage dividers are integrated into setups such as COALA (TU Darmstadt), BECOLA (Michigan State University), COLLAPS (CERN/ISOLDE), and ATLANTIS (Argonne National Laboratory). Monitoring the applied acceleration potential, these dividers allow one to consider and counteract long-term drifts and thereby improving measurement accuracy.

## 1. INTRODUCTION

This paper presents a state-of-the-art concept for an ultra-stable, thermally independent precision voltage divider designed for direct current (DC) voltages up to 60 kV. Notable features of this voltage divider include its minimal voltage dependence, exceptional stability, and resistance to external temperature influences. The innovative aspect is that it is manufactured using 3D printing technology, making it easy for project partners to replicate. The precision voltage divider demonstrates the seamless integration of precision resistors, readily available in the market, building upon the successful outcomes observed in the FutureEnergy 19ENG02 Project and HVDC ENG07 Project [Houtzager.2012].

The precision voltage divider addresses the critical need for accurate voltage measurement in high-voltage applications. Its low voltage dependence ensures reliable performance over a wide voltage range, while its exceptional stability guarantees accurate and consistent measurements over time. In addition, the device is insensitive to external temperature variations, making it an ideal choice for use in a wide range of environmental conditions.

Accurate high-voltage measurements are essential for a manifold of modern high-precision experiments in particle physics, e.g., for determining the neutrino mass by measuring the electron-endpoint energy [Aker.2022, Thümmler.2009], nuclear physics, e.g., for the determination of masses [Wienholtz.2013, Wienholtz.2020] or charge radii [Otten.1989, Krieger.2011], and atomic physics, e.g., in storage ring experiments [Ullmann.2017]. The requirements of the different applications differ: While trap experiments are mostly interested in monitoring and counteracting drifts of the potentials applied to the electrodes [Boehm.2016, Fischer.2021], at storage rings or the neutrino-mass experiment, the absolute voltage needs to be measured to determine the kinetic energy of the particles [Lochmann.2014, Arenz.2018, Meisner.2020]. The divider presented in this work was designed for fast-beam collinear laser spectroscopy [Kaufman.1976, Yang.2023] at energies of up to 60 keV, for which both aspects are of particular importance.

In the relevant experiments, an ion beam is electrostatically accelerated either directly from an ion source or from a beam-preparation trap. A laser beam is then superimposed on the ion beam, and electronic transitions in the atomic shell are probed. To extract the transition frequency, the ion beam velocity and, hence, the acceleration potential must be known accurately. These frequencies are interesting, e.g., for determining absolute charge radii and for astrophysical studies. Therefore, at the collinear laser spectroscopy setups COALA (TU Darmstadt) and BECOLA (Michigan State University), the voltage dividers described in this paper are combined with a digital feedback loop that counteracts long-term drifts of the applied potential. Further copies of the divider are also used in the collinear laser spectroscopy setups COLLAPS (CERN/ISOLDE) and ATLANTIS (Argonne National Laboratory).

In addition, the 3D-printed precision voltage divider not only underlines technological innovation but also facilitates easy replication by project collaborators. The seamless integration of precision resistors and resistance to external temperature effects make this voltage divider a valuable tool for accurate voltage measurements in various scientific applications, contributing to advances in metrology and experimental physics.

## 2. DESIGN OF THE DIVIDER

The following section focuses on the divider's design, with special emphasis on the 3D-printed structure and temperature stabilisation.

## 2.1. Mechanical structure

The main aim of the mechanical design of the divider is to be easy to replicate and to provide the basis for a high precision divider. Therefore, the divider's high voltage arm is designed to be printed using a readily available FDM (Fused Deposition Modelling) 3D printer. The parts are printed with PETG (polyethylene terephthalate modified with glycol) because it is not hygroscopic and, therefore, does not absorb water. There is no need for a support structure while printing, making finishing the parts easier. One module can hold 24 resistors, forming a helix with a distance of 24 mm between each winding, which can be seen in the following figure.

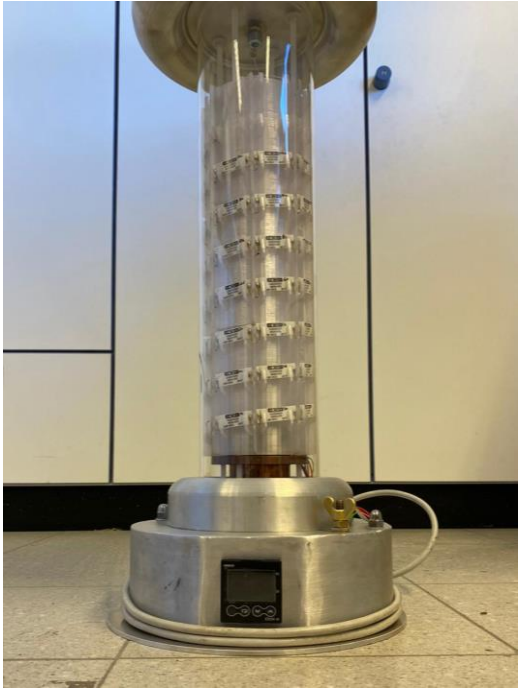

Figure 1. Foto of the finished high voltage divider

Due to the helical arrangement, the band of resistors forms an even electrical field, resulting in a partial-ischarge free operation. The homogenous electrical field distribution is especially important since 3D-printed parts usually have voids. The individual modules are held together with the aid of four nylon threaded rods with a thread size of M6.

These rods clamp the whole assembly against a copper block on which the heating element is mounted, which in turn is screwed to the base of the divider. This base contains the power supply and the temperature control unit, as well as a ground connection and a connection for the measurement output signal. Within the temperature-controlled zone is the low-voltage resistor across which the output signal is measured. A fan circulates the air to ensure uniform temperature throughout the assembly. The fan sits under the copper block and blows heated air through the centre section of the 3D-printed support structure. The air return path is on the outside of this structure, where the resistors are located. A wire mesh is placed between the copper block and the 3D-printed support structure and connected to the ground to shield interference from the fan and discharge any particles that may be circulating in the air stream.

The whole assembly is placed inside an acrylic tube to provide thermal, dust, and moisture shielding. The top of the divider is closed by an aluminium corona ring. This ring ensures a homogeneous electrical field distribution along the entire divider, thus preventing any corona discharges that might otherwise lead to parasitic currents.

## 2.2. Electrical design

The temperature controller is an off-the-shelf OMRON E5CN-HV2M-500 device. The connected PT100 sensor measures the temperature in the airflow inside the divider. A MJH11022G Darlington transistor generates heat and is mounted on a heatsink in the airflow. The temperature controller (E5CN-HV2M-500) controls the transistor with an analog signal (0-10 V) via the 700-$\Omega$ resistor. Depending on the transistor's current amplification factor, the resistance value may need to be adjusted to achieve sufficient current flow in the CE path of the transistor to heat the air.

Three silicon diodes connected in series from the base of the transistor limit the maximum voltage at the base to ground and generate current feedback in conjunction with the emitter resistor of 0.2 $\Omega$. This limits the maximum power dissipation at the transistor and improves the adjustment of the operating point. The emitter resistor also ensures that the feedback minimises the effect of the transistor's thermal drift. The circuit is shown in the following figure.

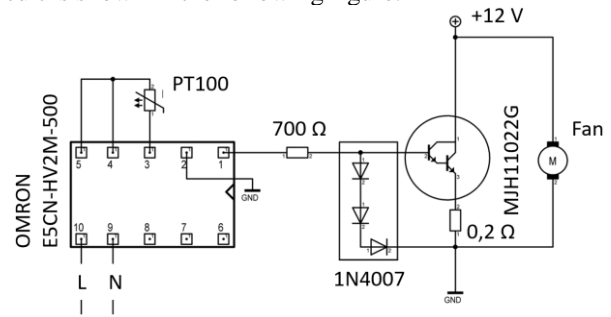

Figure 2. The circuit design of the temperature regulation with an E5CN-HV2M-500 regulator and an MJH11022G Darlington transistor as the heating element.

## 2.3. Measurement circuit design

The high voltage arm of the divider consists of a series of Caddock MDN1475, which are similar to the USFS370 [Caddock] thin film resistors with a Vishay metal foil resistor as a low voltage arm. The high voltage resistors are sold in sets of five, which are then matched in terms of their temperature coefficient to be below 1 $\mu\Omega/\Omega/K$ for the whole set. A resistance value of 10 M$\Omega$ per single resistor and a nominal operating voltage of 1000 V results in a current of 0.1 mA at full voltage. This current has proven to be the optimal choice between reasonable self-heating and negligible effects of leakage currents [Passon.2016]. The low-voltage resistor has a value of 100 k$\Omega$ to give a maximal output voltage of 10 V at the nominal operating voltage. A higher output voltage is not reasonable since most voltmeters on the market have an input impedance of > 10 G$\Omega$ only in the 10-V range or lower. When the 100-V range is used, a voltmeter internal voltage divider will influence the low voltage arm resistor with its own parallel resistance.

In addition, if higher values for the low voltage arm resistance are chosen, the bias current of a voltmeter will become a non-negligible issue [Rietveld.2004]. If lower values for the resistance are chosen, then the offset voltages due to different contact materials will play a major role. Furthermore, interferences will have a greater effect as the signal-to-noise ratio (SNR) decreases.

Additional factors have been considered to further reduce the effect of noise on the measurement signal. An important role is the cable connecting the divider with the precision voltmeter. A shield around the two signal conductors acts as a shield for external electrical fields. A triaxial cable is the best choice if magnetic interferences couple into the cable since the effects will cancel each other out. At first, a shielded cable with two inner conductors was used with PTFE insulation for the presented dividers. The PTFE insulation is superior regarding leakage currents; however, it has the major disadvantage of having a triboelectric effect. The slightest movement of the cable will result in a voltage produced by this effect, which distorts the measurement. Therefore, later, the cables were replaced with Belden 9222 with a true coaxial structure, and the connectors were replaced with Lemo 2S triaxial connectors. The outermost connector acts as the shield, often called the guard. The innermost connector is connected to the high signal of the divider, and the connector in between is the low signal. The following figure shows the connection from the divider on the left via the triaxial cable to the voltmeter illustrated in grey on the right.

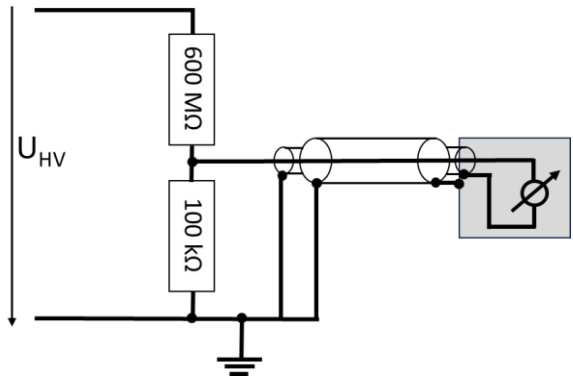

Figure 3. Circuit of the 60 kV high voltage divider with triaxial connector and triaxial Belden 9222 cable

### 3. MEASUREMENT SETUP

In collinear laser spectroscopy experiments, an ion beam is electrostatically accelerated either directly from an ion source or a beam-preparation trap. The ion beam is then superimposed with a laser beam, and electronic transitions in the atomic shell are probed, i.e., an electron is excited if the laser frequency matches the specific transition frequency $f_0$ in the rest frame of the ion, i.e., $f_{ion} = f_0$. However, since the ion beam possesses a kinetic energy of several 10 keV, the ions experience the laser frequency $f_{Laser}$ Doppler shifted

$$f_{ion} = f_{Laser}\, \gamma(1 \pm \beta)$$

with the relativistic factor $\gamma = (1 - \beta^2)^{-1/2}$, the velocity β in units of the speed of light, a negative sign if the laser and ion beam are copropagating and a positive sign if both beams are counter-propagating.

For extracting the transition frequency $f_0$, the ion-beam velocity and, hence, the acceleration potential must be accurately known. Those are of interest, e.g., for the determination of absolute charge radii [Imgram.2023] and astrophysical studies [Berengut.2011], for which accuracies at the 1-MHz level are required. Under typical conditions, this corresponds to a precision of approximately 100 mV. To extract relative charge radii in an isotopic chain or electromagnetic moments, the requirements are relaxed to the 1-V level, but drifts between those relative measurements become critical.

Therefore, two identical copies of the present divider were built to support the laser spectroscopy measurements at CERN/ISOLDE (Collaps [Neugart.2017]) and at Argonne National Laboratory (Atlantis). Those are well suited to monitor drifts of the applied acceleration potential and, hence, drifts of the beam velocity. Determining the absolute beam velocity from voltage measurements is more complicated as, e.g., contact and thermal potentials and field penetration can lead to systematic shifts but can be circumvented by applying laser spectroscopic approaches [König.2021].

### 4. RESULTS AND DISCUSSION

The following section first describes the calibration of the divider at the high-voltage metrology laboratory of the Physikalisch-Technische Bundesanstalt and then the measurements carried out at ISOLDE/CERN and Argonne National Laboratory.

*4.1. Calibration*

The calibration of the divider was carried out in the HVDC metrology laboratory of PTB. Two different measurement procedures are carried out. The first is a stability measurement, and the second is a linearity measurement. The same reference divider MT100 [Marx.2000] is used for both tests, which is well known for its own performance. Type 3458a voltmeters in the 10 V range are used to record the low-voltage signals. These voltmeters are connected to the dividers using Belden 9222 triaxial cables with Lemo triaxial connectors.

To avoid corona discharges, flexible stainless steel tubes with a diameter of 50 mm are used for the high-voltage connection.

The 3458a voltmeters that were used for all measurements are calibrated against a 10 V reference voltage source, corrected for offset voltage, and triggered simultaneously. An integration time of 50 NPLC, equivalent to 1 second, was chosen.

## 4.2. Stability

The purpose of these measurements is to determine the short-term stability of the scale factor of the divider. Therefore, a voltage is increased to the nominal voltage within a few seconds, and the output signal is recorded. The drift of the output signal then depends on the voltmeter, the reference divider, and the divider under test itself. Then, the dividers signal is mainly affected by the self-heating of the resistors and space- or surface charge effects. The self-heating itself can be subdivided into two separate effects. One is the heating of the resistors due to the measurement current of up to 100 µA, which happens typically within several seconds to a few minutes, depending on the resistor topology. The second effect is the heating of the dividers housing, which in this case is counteracted by the temperature regulation and consequently negligible.

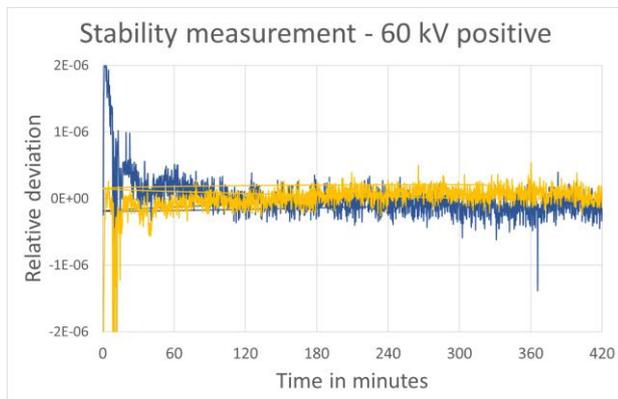

Figure 4. Stability measurement of two of the 60 kV temperature-regulated high voltage dividers at a high voltage of 60 kV positive voltage over a period of 420 minutes (7 hours).

These measurements show a drift of less than 3 µV/V within the first 10 minutes. After that, all measurement points were within 0.5 µV/V. This drift could originate from the dividers or even from the reference measurement system since its best measurement uncertainty is 2 µV/V.

The two dividers presented in this paper were both built in April 2021. A recalibration of one of the dividers at the end of 2023 showed a drift of the scale factor of $27.2 \cdot 10^{-6}$. This drift is quite significant, but the fact that the divider was calibrated a few days after it was built can explain this drift since the resistors had not aged then at all. In addition, a high voltage was applied continuously over a period of two years, which could also lead to this significant drift.

## 4.3. Linearity

A linearity measurement reveals the voltage dependency of the divider. Thus, the voltage range should cover the lowest voltage for which the divider is used up to the full nominal voltage. Before the measurement, an offset voltage determination must be done for the entire measurement chain. Therefore, the high voltage terminal of the divider gets connected to ground, and the output voltage is measured in the voltage range that is also used later on. From this measurement, a voltage coefficient of $1 \cdot 10^{-7}$ kV$^{-1}$ was derived, which is expected due to the thin-film character of the resistors [3].

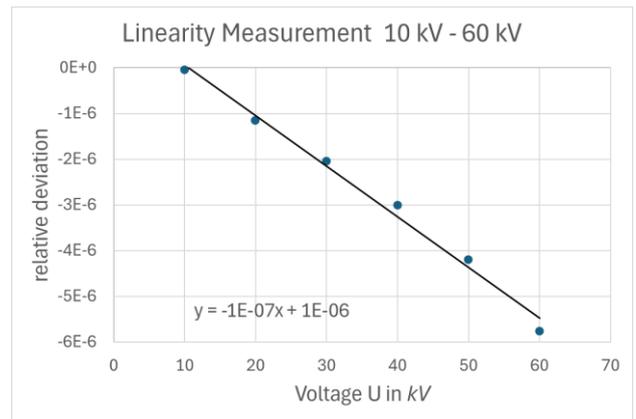

Figure 5. Linearity measurement of the 60-kV temperature-regulated high-voltage divider from 10 kV to 60 kV positive voltage.

## 4.4. Laser spectroscopy results

Two copies of the present divider were built for the laser spectroscopy experiments at ISOLDE/CERN and Argonne National Laboratory and installed in 2021. As the used acceleration potentials are smaller than 60 kV, the dividers were designed with a voltage divider ratio of 6000:1, having a total resistance of 600 MΩ, as described above. The dividers were successfully operated during several measurement campaigns, and exemplary results are shown in Fig. 6, in which the measured acceleration voltage is plotted throughout the experiment. In (a) neutron-rich Ru (24 kV, 3 days) was measured at ANL, in (b) neutron-rich Ca (20 kV, 3 days) using the General Purpose Separator (GPS) at ISOLDE, and in (c) neutron-deficient Tl (50 kV, 12 days) using the High-Resolution Separator (HRS) at ISOLDE. Figure 6 depicts the voltage relative to the mean voltage over time.

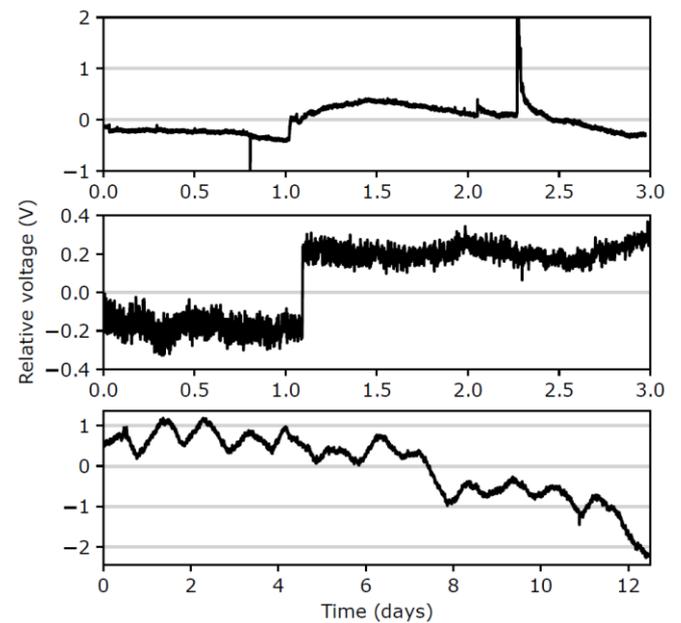

Figure 6. Applied acceleration voltage relative to the overall average at (a) the Ru experiment (2022) at ANL at a total voltage of 24 kV, (b) the Ca experiment (2023) at ISOLDE using the General Purpose Separator (20 kV) and (c) the Tl experiment (2023) at ISOLDE using the High-Resolution Separator (50 kV).

Before the introduction of high-precision, high-voltage dividers to these experiments, the acceleration voltage in the experiments was assumed constant or exhibiting a linear trend between laser-spectroscopic reference measurements that were only performed every few hours. This assumption, although reasonable since high-voltage power supplies with the highest available voltage stability are used, however, misses spikes and jumps as, e.g., observed from the ANL power supply. Besides those spikes and jumps, the ANL power supply showed drifts of up to 100 mV/h. Contrary, the GPS power supply shows a high long-term stability, except for one sudden jump by 400 mV, but has a 200-mVpp noise. As the laser spectroscopic measurements are relatively slow, a fast noise is less critical than a long-term drift or jumps but increases the linewidth and, hence, limits the resolution. The HRS yielded the worst performance with day-night drifts of 1 V and a total drift of 3 V throughout the experiment.

As the aimed precision of the laser spectroscopic measurements is of the order of 1 MHz and a 1-V drift causes frequency changes of typically 5-30 MHz, depending on the mass and optical transition of the investigated ions and the total acceleration potential itself, voltage stability < 100 mV during the measurement of one exotic isotope, that typically takes 1-4 h, is targeted. Only the measurements with the present voltage dividers revealed that this is not the case for the power supplies in use. Interpolating the high-voltage from two reference measurements is thus to be performed carefully since it is only applicable if no voltage jumps from the supplies are observed.

The new dividers allow us to measure the acceleration potential in situ and either correct the data in the analysis procedure or stabilize the high voltage. The latter was demonstrated for collinear laser spectroscopy experiments at TU Darmstadt and at Michigan State University [König.2024] and was recently also applied at ANL. Figure 6 shows the corresponding laser spectroscopy measurement, in which the acceleration potential was stabilized with a 5-V DAC on top of a high-voltage platform to the divider reading. The extracted resonant laser frequencies that were obtained over 24 h agree well with one another and indicate no further drifts.

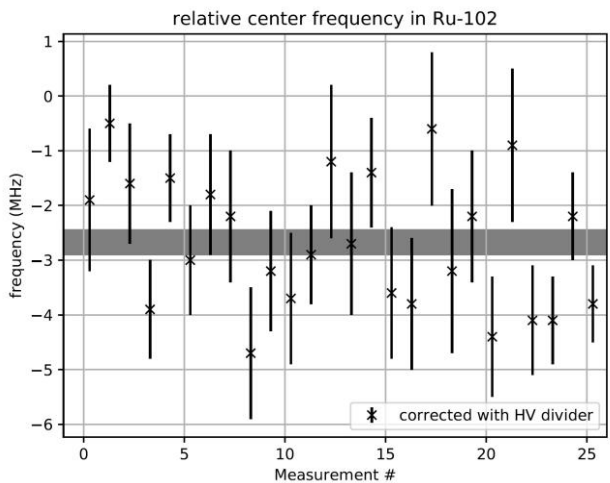

Figure 7. Resonant laser frequencies of $^{102}$Ru relative to the overall average from a measurement series in which the acceleration voltage was stabilized to the voltage divider output. The total measurement time was 24 h. The shaded area corresponds to the standard deviation of the mean of all measurements.

This demonstrates that the described new voltage dividers are crucial tools for precise collinear laser spectroscopy experiments and can contribute to the accurate determination of nuclear ground state properties.

## 5. CONCLUSIONS

In conclusion, developing the precision voltage divider described in this paper represents a significant advancement in high-voltage metrology, particularly for applications in experimental physics such as fast-beam collinear laser spectroscopy. The divider's unique design, incorporating 3D printing technology and precise temperature stabilization, ensures exceptional stability and reliability in voltage measurements up to 60 kV.

Key outcomes of this work include:

1. **Innovative Design and Fabrication:** 3D printing technology in the divider's construction enables precise replication and ensures a robust, homogeneous electrical field distribution critical for high-voltage operations.
2. **Thermal Stability:** Integrating temperature regulation mechanisms guarantees minimal temperature dependence, essential for accurate and consistent voltage measurements over time and in varying environmental conditions.
3. **Experimental Applications:** The divider has been successfully deployed in critical experiments at leading research institutions like CERN/ISOLDE and Argonne National Laboratory, where precise voltage measurements are pivotal for determining nuclear properties and studying nuclear physics phenomena.
4. **Measurement Performance:** Calibration and performance testing demonstrate excellent stability, with minimal drift observed over extended measurement periods. This stability is essential for achieving the precision required in modern experimental physics.
5. **Impact on Laser Spectroscopy:** The divider's deployment in collinear laser spectroscopy experiments has shown significant improvements in stabilizing acceleration potentials, crucial for maintaining the accuracy and resolution of such experiments.

In summary, the precision voltage divider presented in this work addresses the critical need for accurate voltage measurements in high-voltage applications and paves the way for enhanced experimental capabilities in fields ranging from particle physics to nuclear and atomic sciences. Its successful integration into complex experimental setups underscores its importance as a fundamental tool in advancing scientific understanding and discovery.

## ACKNOWLEDGMENTS

We would like to express our gratitude to Finn Köhler for his invaluable contribution to this research. His meticulous efforts in conducting the measurements were instrumental to the success of this paper.